\documentclass[twocolumn,showpacs,preprintnumbers,amsmath,amssymb]{revtex4}

\usepackage{graphicx}
\usepackage{dcolumn}
\usepackage{bm}
\usepackage{graphics}
\usepackage{latexsym}
\usepackage{graphics}
\usepackage{dcolumn}
\usepackage{epsfig}
\usepackage{amssymb} 
\usepackage{amsmath} 
\usepackage{rotating} 
\usepackage{rotate} 
\usepackage{color} 
\usepackage{graphicx}
\usepackage{amssymb}
\newcommand{\gammap}{\dot{\gamma}}

\newcommand{\sta}{R_{\scriptscriptstyle 2}}
\newcommand{\rot}{R_{\scriptscriptstyle 1}}

\begin{document}
\title{Shear-banding in a lyotropic lamellar phase\\ 
Part 2: Temporal fluctuations}
\author{Jean-Baptiste Salmon}
\email{salmon@crpp-bordeaux.cnrs.fr}
\author{S\'ebastien Manneville}
\author{Annie Colin}
\affiliation{Centre de Recherche Paul Pascal, Avenue Schweitzer, 33600 PESSAC, FRANCE}
\date{\today}
\begin{abstract}
We analyze the temporal fluctuations of the flow field associated to  
a shear-induced transition in a lyotropic lamellar phase: 
the {\it layering transition} of the {\it onion texture}. In the first 
part of this work [Salmon {\it et al.}, submitted to Phys. Rev. E], we have evidenced banded flows
at the onset of this shear-induced transition which are well accounted for by  
the classical picture of {\it shear-banding}. In the present paper, we focus on the temporal fluctuations
of the flow field recorded in the coexistence domain. These striking dynamics are very slow (100--1000~s)
and cannot be due to external mechanical noise. 
Using velocimetry coupled to structural measurements, we show that these
fluctuations are due to a motion of the interface separating the two differently sheared bands.
Such a motion seems to be governed  by the fluctuations of $\sigma^\star$, the local stress 
at the interface between the two bands.
Our results thus provide more evidence for the relevance of the classical mechanical approach of shear-banding  
even if the mechanism leading to the fluctuations of $\sigma^\star$ remains unclear.
\end{abstract}
\maketitle

\section{Introduction \label{sec:introduction}}

Emulsions, shampoos, and paints are example of everyday life complex fluids. 
A {\it mesoscopic} scale located between the molecular size and the sample size is 
one of the most important features of these complex materials. In the case of an oil-in-water emulsion for instance, 
this length scale is the diameter of the oil droplets ranging from 100~nm to a few microns. 
During the last decade, many experimental and theoretical works have 
been devoted to the understanding of the effect of 
shear on these fluids \cite{Larson:99,Edimbourg:00}. 
A robust fact has emerged: the flow modifies the structure 
of the fluid, which in turn changes the flow field. 

In some cases, this strong coupling between flow and structure 
may induce the nucleation of a new phase or even create new structures or {\it textures} that do not exist at rest.
In wormlike micelles for instance, shear may induce a nematic phase \cite{Berret:94,Schmitt:94}, whereas 
in membrane phases, shear creates new textural organizations \cite{Diat:93,Roux:93}.   
Such Shear-Induced Structures (SIS) usually appear above a critical shear rate $\gammap$ or shear stress $\sigma$.
For some range of controlled $\gammap$ (or $\sigma$),  the SIS coexists with 
the unmodified structure, 
and then progressively fills the flow as $\gammap$ (or $\sigma$) 
is further increased. Usually, the SIS flows very differently 
from the initial structure and large variations of the {\it effective} viscosity $\eta\widehat{=}\sigma/\gammap$ 
are recorded near the shear-induced transition. Since the two structures have different viscosities, 
inhomogeneous flows or {\it banded flows}, are expected.
Such an experimental picture is referred to in the literature as {\it shear-banding} \cite{Larson:99,Edimbourg:00}.  
Shear-banding scenarii have been reported in a wide class of materials 
using both structural and rheological measurements
\cite{Cappelaere:97,Hu_1:98,Eiser:00,Panizza:95,Roux:93,Ramos:00}. 
Moreover, a few experiments based on local velocimetry have reported the existence of such inhomogeneous flows 
where bands flow at different shear rates \cite{Debregeas:01,Coussot:02,Pignon:96,Mair:96,Britton:97,Fischer:01,Hu_2:98,Salmon:03_4}.
These experiments mainly concern glassy materials and wormlike micellar systems. 

In the first part of this work \cite{Salmon:03_6}, we have started an exhaustive study of a 
shear-induced transition in a lyotropic lamellar phase composed of Sodium Dodecyl
Sulphate (6.5\%~wt.) and Octanol (7.8\%~wt.) in Brine at 20~g.L$^{-1}$.  
Since the pioneering work of Roux and coworkers \cite{Diat:93,Roux:93}, this system has been widely studied
\cite{Diat:95,Sierro:97,Leng:01}.  
For $\gammap\gtrsim 1$~s$^{-1}$, the membranes are wrapped
into multilamellar vesicles called {\it onions}, that form a monodisperse disordered close-compact assembly.
The characteristic size of the onions is a few microns. 
At higher shear rates ($\gammap\geq 15$~s$^{-1}$), onions get a long range hexagonal order and are organized on 
layers sliding onto each other.
This shear-induced ordering transition is referred to in the literature as the {\it layering} transition \cite{Ackerson:88}.  
Using local velocimetry, coupled to structural and rheological measurements \cite{Salmon:03_6}, we have
shown that above a critical shear rate, a highly sheared band, most probably composed of the ordered texture, is nucleated
at the rotor of the Couette cell. This band coexists with the viscous disordered state thus leading to a banded flow.
When $\gammap$ is further increased, the highly sheared band grows up to finally invade the whole gap of the Couette cell. 
The classical picture of shear-banding thus holds for the layering transition. 

However, one point remains very striking. Indeed, the flow field presents 
large temporal fluctuations of the local velocity in the coexistence domain.
Such fluctuations reach up to 20\% at the interface separating the two differently sheared bands.   
This phenomenon seems quite general in shear-banded flows since such temporal fluctuations 
have also been detected in some wormlike micellar systems using local velocimetry \cite{Britton:97,Fischer:01}.
On a more general ground, it has been shown recently that some complex fluids may also exhibit some dynamics
when submitted to a shear flow in the vicinity of shear-induced transitions 
\cite{Hu_1:98,Pignon:96,Bandyopadhyay:01,Lootens:03,Wheeler:98}.
In particular, in our lamellar phase, the effective viscosity $\eta=\sigma/\gammap$ 
presents aperiodic or relaxational oscillations under controlled $\sigma$, 
suggesting deterministic behaviors in a very narrow range of parameters \cite{Wunenburger:01,Salmon:02}.
For all these dynamical behaviors, referred to as {\it rheochaos}, 
it seems that spatial degrees of freedom play a relevant role \cite{Pignon:96,Bandyopadhyay:01,Salmon:02}. 
On the experimental point of view, it is important to determine whether the spatial organization
of the flow field near such transitions is responsible for the observed dynamics.
Such experimental data may then be very helpful to unveil the most relevant ingredients of 
the existing theoretical models \cite{Head:01,Cates:02,Picard:02}.   

In this second part of the article, we present a detailed study of the temporal fluctuations
in our system. In the next section, after recalling briefly 
the experimental setup, we analyze 
the temporal fluctuations of the global measurements and of some time series of the local velocity. 
In Sec.~\ref{sec:efadotb}, we show that these temporal fluctuations
are due to a motion of the interface separating the bands 
inside the gap. We then come back in Sec.~\ref{auasma} to the mechanical
approach used to describe the shear-banding scenario in Ref.~\cite{Salmon:03_6}. 
Assuming that the stress $\sigma^\star$, at which the interface between the disordered and the ordered textures is stable, 
fluctuates in time, we are able to describe the dynamics of the flow field recorded under 
controlled shear rate but also under controlled shear stress.

\section{Global and local fluctuations in the coexistence domain \label{sec:gfalf}}

\subsection{Experimental setup}

\begin{figure}[ht!]
\begin{center}
\scalebox{1.0}{\includegraphics{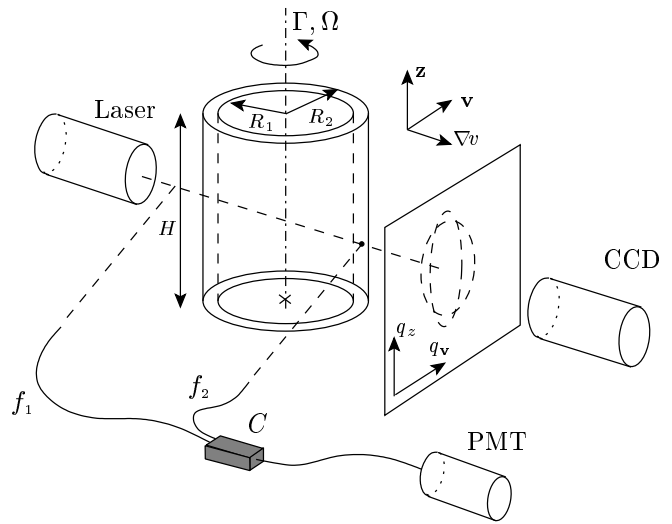}}
\end{center}
\caption{
Experimental setup. A thermostated plate (not shown) on top of the cell 
avoids evaporation. The geometry of the Couette cell is:
$H=30$,  $\rot=24$, and $\sta=25$~mm, leading to a gap width $e=\sta-\rot=1$~mm.
$f_i$ are single mode fibers and $C$ a device coupling the two fibers. PMT denotes a photomultiplier tube and 
CCD the Charge-Coupled Device camera.
The scattering volume is denoted by $\bullet$ and its characteristic size is about 50~$\mu$m.
\label{setup2}}
\end{figure}

The experimental setup has been fully described in the first part of this work \cite{Salmon:03_6}. 
It allows us to perform rheological experiments, local velocity measurements and to
determine the structure of the fluid at the same time. 
Figure~\ref{setup2} sketches the main features of this setup.
A controlled-stress rheometer 
imposes a constant torque $\Gamma$ on the axis of a Couette cell and
records its rotation speed $\Omega$ in real time. From these two {\it global} quantities, the rheometer
indicates the following stress $\sigma$ and shear rate $\gammap$:
\begin{eqnarray}
\sigma &=& \frac{R_{\scriptscriptstyle 1}^2+R_{\scriptscriptstyle 2}^2}
{4\pi H R_{\scriptscriptstyle 1}^2 R_{\scriptscriptstyle 2}^2}\,\Gamma\,,\label{e.sigmarheo} \\ 
\gammap &=& \frac{R_{\scriptscriptstyle 1}^2+R_{\scriptscriptstyle 2}^2}
{R_{\scriptscriptstyle 2}^2-R_{\scriptscriptstyle 1}^2}\,\Omega\,,
\label{e.gammarheo}
\end{eqnarray}     
where $H$, $\rot$, and $\sta$ are the geometrical characteristics of the Couette cell (see Fig.~\ref{setup2}).
The values ($\sigma$,$\gammap$) given by the rheometer represent the spatial averages
of the {\it local} stress $\sigma(r)$ and of the {\it local} shear rate $\gammap(r)$ over the gap of the Couette cell,
in the case of a Newtonian fluid ($r$ is the radial position in the flow). Temperature is controlled 
using a water circulation around the Couette cell. Since the system under study is very sensitive to the
exact fractions of its components, great care has been taken to minimize evaporation using 
a thermostated plate on top of the cell.  

A laser beam crosses the transparent Couette cell along the velocity gradient direction $\nabla\!v$.
The corresponding diffraction patterns are collected by a Charge-Coupled Device camera on a screen
in the ($q_{\mathbf{v}}$,$q_z$) plane.     
Dynamic Light Scattering (DLS)  
experiments in the heterodyne geometry are performed to measure the local velocity.
Such a  method along with its spatial and temporal resolutions, has been described at length
in Ref.~\cite{Salmon:03_2}.
The light scattered by a small volume of the sample (denoted by $\bullet$ in Fig.~\ref{setup2})
is collected using a single mode fiber.
The scattered electric field is then mixed with the incident beam using an optical device coupling 
the two fibers. The Doppler frequency shift associated to the flow field inside the scattering volume 
is computed from the correlation function measured using an electronic correlator connected to a photomultiplier tube 
(see Fig.~\ref{setup2}).
From the Doppler frequency shift, the mean value of the velocity inside the scattering volume is then
obtained using a careful calibration of the setup \cite{Salmon:03_2}.
The characteristic size of the scattering volume is about 50~$\mu$m.  
To obtain velocity profiles in the gap of the Couette cell, the rheometer 
is moved along the direction of shear $\nabla\!v$ by steps of $30~\mu$m.
Note that the measurement of the local velocity at a specific position in the flow lasts about 3--5~s.
A full velocity profile is then obtained in about 2--3~min.

\subsection{Temporal fluctuations of the global rheology}

\begin{figure}[ht!]
\begin{center}
\scalebox{1.0}{\includegraphics{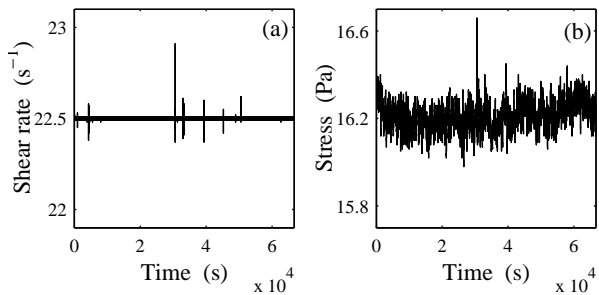}}
\end{center}
\caption{
$T=30^\circ$C.
(a) Time series of the controlled shear rate $\gammap(t)$.
(b) Corresponding temporal response of the measured shear stress $\sigma(t)$.}
\label{global_rheology}
\end{figure}

The experimental procedure used to prepare the onion texture is the following. The
temperature is set at $T=30^\circ$C. 
A constant shear rate $\gammap=5$~s$^{-1}$ is applied for 7200~s. This first step allows 
us to start the experiment with a well-defined stationary state of disordered onions. 
The shear rate is then increased from $\gammap=5$ to $20$~s$^{-1}$ by applying different steps
lasting 5400~s. 
The increment between two different steps is 5~s$^{-1}$. Finally, a constant shear rate of 
$\gammap=22.5$~s$^{-1}$ is applied. Figure~\ref{global_rheology} presents the evolution 
of the controlled $\gammap$ and that of the
measured $\sigma$ during this last step. Let us recall that our rheometer 
imposes a torque $\Gamma$. A constant rotation speed $\Omega$ and thus a constant shear rate $\gammap$ is  
obtained by a computer-controlled feedback loop on the applied $\Gamma$. As shown in 
Fig.~\ref{global_rheology}(a), 
this procedure is very successful since the temporal fluctuations of the controlled $\gammap$ 
are less than $0.05\%$. However, some variations greater than 
0.1~s$^{-1}$ are sometimes recorded but they are very rare and probably induced by exterior mechanical noise. 
The measured stress $\sigma(t)$ is almost stationary: the temporal fluctuations of  
$\sigma(t)$ are about $0.4\%$ [see Fig.~\ref{global_rheology}(b)].  
The non-equality between the
fluctuations of the controlled $\gammap$ and the measured $\sigma(t)$ 
raises important questions: such a difference might result from bad 
filling conditions of the Couette cell or may hide more fundamental points
in the rheological analysis. 

\begin{figure}[ht!]
\begin{center}
\scalebox{1.0}{\includegraphics{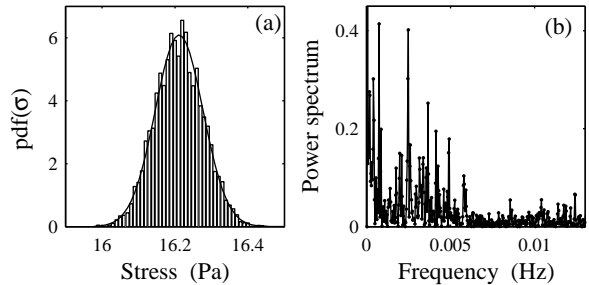}}
\end{center}
\caption{(a) Probability density function of the time series $\sigma(t)$ displayed in 
Fig.~\ref{global_rheology}(b). The continuous line is a Gaussian distribution with mean
16.21~Pa and with standard deviation 0.065~Pa. (b) Power spectrum of the time series $\sigma(t)$ 
displayed  in Fig.~\ref{global_rheology}(b).}
\label{pdf_tf_stress}
\end{figure}

To proceed further in this analysis, we plot the probability
density function of $\sigma$ in Fig.~\ref{pdf_tf_stress}(a).
$p(\sigma)$ is very well fitted by a
Gaussian distribution centered around the mean value $<\!\sigma\!>=16.21$~Pa and characterized by a standard 
deviation of 0.065~Pa.
In Fig.~\ref{pdf_tf_stress}(b), we try to evidence some characteristic times by computing the frequency
power spectrum of the signal.  To minimize side effects, we first 
multiply the time series $\sigma(t)$ by a Hanning function. The frequency power spectrum 
presents a bump around 0.002~Hz corresponding to
characteristic times of the order of 500~s. These time scales are short enough to be
statistically well described by our measurements, since our signal was 
recorded during $6.10^4$~s with a sample rate of 1~s. 
Let us note however that the large peaks at very low 
frequency ($\approx 1.10^{-4}$~Hz) are certainly due to a lack of statistics. 

These above time scales ranging between 100 and 1000~s are very long when compared to 
the period of rotation of the Couette cell ($\approx 5$~s). 
They are reminiscent of the long time scales of the shear rate oscillations 
observed in the same system under controlled stress for temperatures $T\geq T_c=27^\circ$C 
\cite{Wunenburger:01,Salmon:02}. Thus, the 
differences between the amplitude of the fluctuations of $\gammap$ and of $\sigma$ are 
probably not due to a bad filling of the Couette cell but rather hide an important physical point.

\subsection{Temporal fluctuations of the local velocity}

\subsubsection{Fluctuating velocity profile}

\begin{figure}[ht!]
\begin{center}
\scalebox{1.0}{\includegraphics{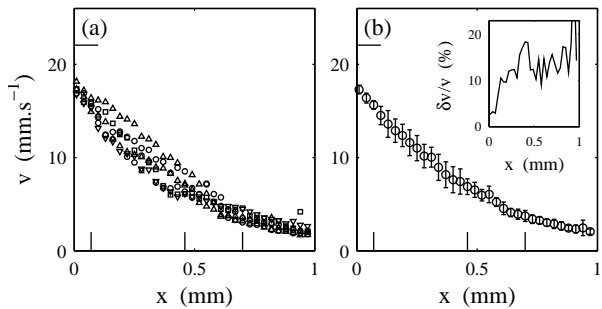}}
\end{center}
\caption{
(a) 6 velocity profiles obtained simultaneously to the rheological data
displayed in Fig.~\ref{global_rheology} ($\gammap=22.5$~s$^{-1}$ and $T=30^\circ$C). 
(b) Corresponding time-averaged velocity profile. The errorbars are the standard deviation inferred
from the 6 profiles displayed in (a). Inset: temporal fluctuations (i.e. standard deviation)
of the local velocity vs. the position $x$. The horizontal lines indicate the imposed rotor velocity
$v_{\scriptscriptstyle 0}$.
The vertical lines indicate the positions $x_{\scriptscriptstyle 1}=0.07$,
$x_{\scriptscriptstyle 2}=0.46$, and $x_{\scriptscriptstyle 3}=0.70$~mm where the 
time series $v(x_i,t)$ displayed in Figs.~\ref{locale_vx2} and \ref{dissym} are measured.}
\label{moyenne_6profils}
\end{figure}

Let us now turn to the local velocimetry measurements.
Figure~\ref{moyenne_6profils}(a) presents various velocity profiles 
obtained under the same applied shear rate of $\gammap=22.5$~s$^{-1}$.
We recall that it takes about 3~min to record a full velocity profile. 
These profiles clearly differ from each other and large variations of the velocity are observed. 
These variations are meaningful and greatly exceed the resolution of our experimental setup.
If we decide, as in the first part of this paper \cite{Salmon:03_6}, 
to average these measurements and to calculate their standard deviations,
we may define errorbars on the time-averaged velocity profile that correspond
to the amplitudes of the temporal fluctuations. 

The time-averaged profile reveals three major issues [see Fig.~\ref{moyenne_6profils}(b)]. 
(i) Sliding occurs: the velocity of the fluid does not vanish at 
the stator and does not reach its expected value at the rotor.
(ii) Moreover, the sliding velocities $v_{si}$, defined as the difference
between the fluid velocity near wall number $i$ and the velocity of wall number 
$i$, are almost constant.
The lubricating layers responsible for the observed sliding, 
do not present any dynamics at the precision of our setup. 
(iii) Two bands with different shear rates coexist in the gap of the Couette cell. 
The highly sheared band is located near the rotor and is about 0.5~mm thick. 

Finally, as shown in the inset of Fig.~\ref{moyenne_6profils}(b),
the amplitude of the relative temporal fluctuations $\delta v/v$ varies with the position in the gap and seems to reach a maximum 
at $x\approx 0.5$~mm where lies the interface between the two differently sheared bands. 
Note that $\delta v/v$ seems to diverge at $x\rightarrow0$: this is due to the small values of 
the velocities near the stator and to the
intrinsic uncertainty of our setup ($\approx 5\%$). 

\subsubsection{Fluctuating time series $v(x,t)$}

\begin{figure}[ht!]
\begin{center}
\scalebox{1.0}{\includegraphics{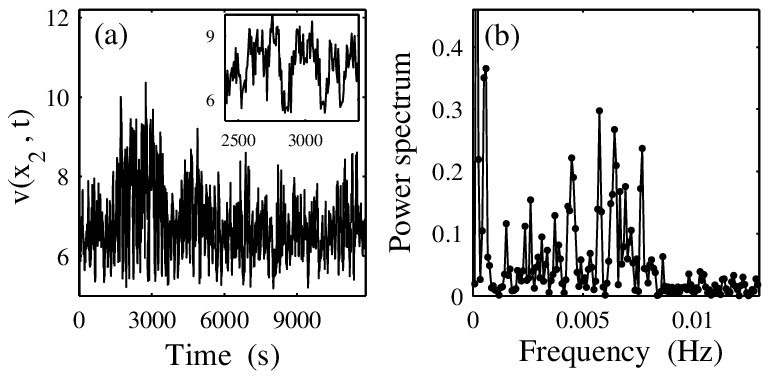}}
\end{center}
\caption{(a) Time series $v(x_{\scriptscriptstyle 2},t)$ measured 
at $x_{\scriptscriptstyle 2}=0.46$~mm, simultaneously to the data
presented in Figs.~\ref{global_rheology} and \ref{moyenne_6profils}.
Inset: blow-up of the dynamics. (b) Correponding power spectrum.}
\label{locale_vx2}
\end{figure}

The variations recorded above refer to the amplitude of the temporal fluctuations. 
To analyze them in more details, we record the evolution of the local velocity at a given position
in the flow field, by measuring successively several correlation functions accumulated over 3~s. 
We thus obtain the evolution of the velocity with a temporal sample rate of 3~s. This 
sample rate is short enough to follow the slow dynamics described above and long 
enough to measure precisely the velocity \cite{Salmon:03_2}. Figure~\ref{locale_vx2}(a) presents the measurement performed in the
middle of the gap at $x_{\scriptscriptstyle 2}=0.46$~mm. 
The signal was recorded during 11500~s.
The inset clearly reveals fluctuations of the local velocity with an amplitude of about $4$~mm.s$^{-1}$ and 
characteristic times ranging from 100 to 500~s. 
The power spectrum displayed in Fig.~\ref{locale_vx2}(b), points out the same feature: 
it presents a bump centered between 
0.002 and 0.01~Hz.
These measurements show that the time series of the local velocity presents large fluctuations that involve 
long time scales. In order to check whether these behaviors occur homogeneously in the entire gap of the Couette cell, we 
have repeated the same measurement for three positions in the gap $x_{\scriptscriptstyle 1}=0.07$, 
$x_{\scriptscriptstyle 2}=0.46$, and $x_{\scriptscriptstyle 3}=0.70$~mm.

\begin{figure}[ht!]
\begin{center}
\scalebox{1.0}{\includegraphics{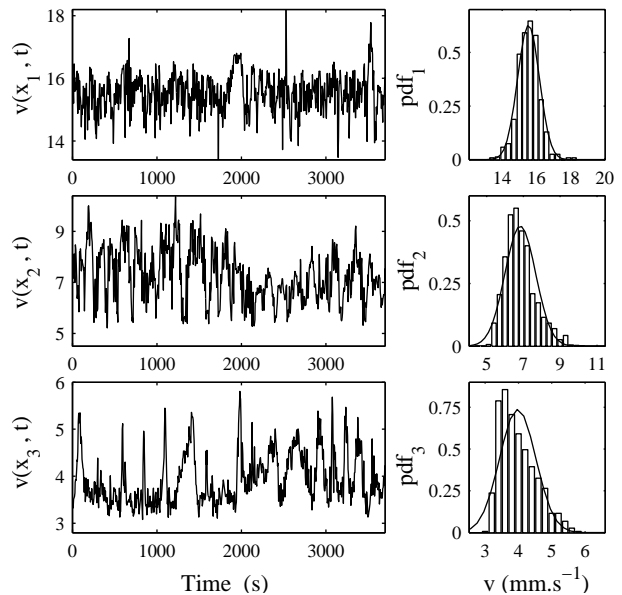}}
\end{center}
\caption{Time series $v(x_i,t)$ and corresponding probability density functions obtained 
simultaneously to the data displayed in Figs.~\ref{global_rheology} and \ref{moyenne_6profils}.
$x_{\scriptscriptstyle 1}=0.07$, $x_{\scriptscriptstyle 2}=0.46$,
and $x_{\scriptscriptstyle 3}=0.70$~mm.
The continuous lines are the Gaussian distributions with the means and the standard deviations of
the time series $v(x_i,t)$.}
 \label{dissym}
\end{figure}

Figure~\ref{dissym} reports these measurements and 
presents the corresponding probability density functions $p(v)$ 
of the local velocities. \\

\noindent 
(i) 
Close to the rotor at $x_{\scriptscriptstyle 1}=0.07$~mm, the velocity does not fluctuate much: 
the relative variation is equal to 4\%. This value is of the same order of magnitude as the 
uncertainties on the measurement. It is thus rather difficult to analyze precisely this data set. However, 
temporal fluctuations on about 100~s are clearly present. 
A Gaussian distribution centered around 15.5~mm.s$^{-1}$ and characterized by
a standard deviation of 0.6~mm.s$^{-1}$ nicely fits the experimental distribution of the local velocity.\\
\noindent (ii) 
In the middle of the gap at $x_{\scriptscriptstyle 2}=0.46$~mm, the fluctuations are much larger and 
reach 12\%. Moreover, the fluctuations are not symmetric: more events occur at high velocity
than at low velocity. The distribution is thus not Gaussian.\\ 
\noindent (iii)
Near the stator at $x_{\scriptscriptstyle 3}=0.70$~mm, this behavior is even more pronounced: the velocity 
seems to fluctuate 
between two states. Most of the time, the velocity remains equal to about 3.5~mm.s$^{-1}$, 
but sometimes the velocity increases rapidely to 5--6 mm.s$^{-1}$. 
The most probable velocity is now clearly smaller than the mean velocity.
The probability density function is highly dissymmetric and 
shows a long tail of rare events at high velocities

These measurements show that (i) long time scales ranging between 100 and 1000~s 
are involved in the dynamics of the local velocities;  (ii)
the amplitude of the fluctuations varies in the gap and reaches a maximum at the interface between the bands; 
(iii) the probability density functions are Gaussian near 
the rotor and become highly dissymmetric near the interface in the weakly sheared band.

\section{Evidence for a motion of the interface between the two bands \label{sec:efadotb}}

\subsection{Local velocimetry measurements}

To proceed deeper into the analysis, a dynamical picture of the flow in the entire gap would be
very helpful. Unfortunately, our experimental setup does not provide an excellent temporal 
resolution: about 2--3~min are necessary to obtain a full velocity profile by moving the rheometer along
the velocity gradient direction $\nabla\!v$. This time is comparable to the time scales involved in the 
temporal fluctuations. Thus, the velocity profile is not {\it frozen} during the measurement. 

In the first part of this article, we have analyzed the time-averaged velocity profiles
obtained using a careful protocol to study the layering transition (see Ref.~\cite{Salmon:03_6}).
Keeping in mind the time needed to obtain a full profile ($\approx 3$~min), we now
focus on {\it individual} profiles, i.e. without averaging. 

\begin{figure}[ht!]
\begin{center}
\scalebox{1.0}{\includegraphics{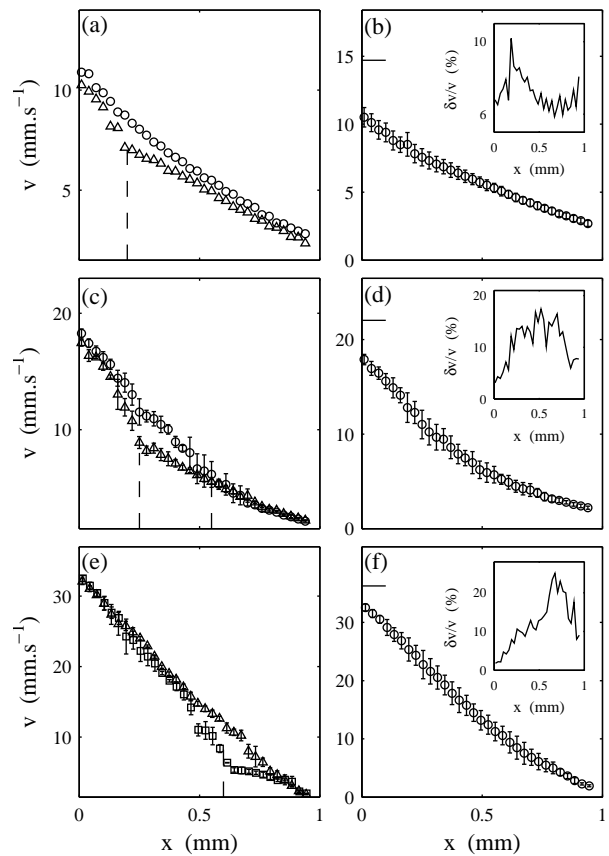}}
\end{center}
\caption{
Experiments at $T=30^\circ$C and under controlled shear rate.
(a) Two {\it individual} velocity profiles obtained at $\gammap=15$~s$^{-1}$.
The vertical dashed line lies at $x=0.20$~mm.
(b) Velocity profile averaged over 13 consecutive measurements at $\gammap=15$~s$^{-1}$.
(c) Two {\it averaged groups} of velocity profiles obtained at $\gammap=22.5$~s$^{-1}$.
The vertical dashed lines lie at $x=0.25$ and $x=0.55$~mm.
(d) Velocity profile averaged over 13 consecutive measurements at $\gammap=22.5$~s$^{-1}$.
(e) Two {\it averaged groups} of velocity profiles obtained at $\gammap=37$~s$^{-1}$.
The vertical dashed line lies at $x=0.60$~mm.
(f) Velocity profile averaged over 20 consecutive measurements at $\gammap=37$~s$^{-1}$.
(b), (d), and (f): the horizontal lines indicate the rotor velocity $v_{\scriptscriptstyle 0}$.
Insets: temporal fluctuations of the local velocity inferred from the standard deviation of the 
different measurements vs. $x$.}
\label{bande_mouvement}
\end{figure}

Figure~\ref{bande_mouvement}(a) presents two individual profiles 
obtained at $\gammap=15$~s$^{-1}$ and the time-averaged profile calculated over 13 successive measurements 
is shown in Fig.~\ref{bande_mouvement}(b).
For this applied shear rate, the time-averaged velocity profile is almost linear. Sliding is observed both
at the rotor and at the stator. 
The inset of Fig.~\ref{bande_mouvement}(b) presents the relative fluctuations $\delta v/v$ vs. the position $x$. 
$\delta v/v$ reaches a maximum of 10\% at 
$x\approx 200~\mu$m. 
Moreover, among the 13 individual profiles, we can distinguish two kinds of measurements: some are linear while 
others present bands supporting different local shear rates [see Fig.~\ref{bande_mouvement}(a)]. 
In this last case, the thickness of the highly sheared band is roughly equal to $200~\mu$m. 
It seems that these two families of profiles correspond to {\it frozen} pictures of 
nearly stationary flow fields.

At this stage, an interpretation of the temporal fluctuations 
begins to emerge. These fluctuations could be correlated to the motion of the interface between the
two bands. If the position of the interface moves, then different kinds of velocity profiles will 
be measured at this low shear rate: linear ones when the highly sheared band is absent, and shear-banded
ones when the two textures coexist. Fluctuations of the position of the interface separating the two bands 
will also lead to a spatial localization of the fluctuations of the local velocity. 
This is indeed the case as can be seen in the inset of Fig.~\ref{bande_mouvement}(b).

In order to further check this idea, we performed the same experiments at higher shear rates. Since the temporal 
fluctuations are more important, frozen pictures of the flow field 
are more difficult to capture. However, some profiles reveal 
the same results as for $\gammap=15$~s$^{-1}$. 
Figure~\ref{bande_mouvement}(c) shows two different velocity profiles averaged over 
some well-chosen profiles among 13 successive measurements at $\gammap=22.5$~s$^{-1}$.  
In both cases, two differently sheared bands coexist in the gap.
However, the position of the interface varies a lot. For the first group of velocity profiles, 
the interface lies at $x\approx0.25$~mm and it has moved to $x\approx0.55$~mm for the second group. 
In both cases, the velocity of the fluid near the rotor is nearly constant.  
This implies that the shear rate in the highly sheared band has changed from 35 to 23~s$^{-1}$.
Note that the shear rate in the weakly sheared band does not seem to vary.  
On average, the interface between the two textures lies at 
$x\approx0.40$~mm and the amplitude of the fluctuations is 
maximal at this position [see Fig.~\ref{bande_mouvement}(d)]. 

The same conclusions may be drawn by looking at two groups of profiles obtained at
$\gammap=37$~s$^{-1}$ and displayed in Fig.~\ref{bande_mouvement}(e).
In this case, the highly sheared band has invaded the gap on a 
{\it time-averaged} point of view since the 
velocity profile averaged over 20 successive measurements is almost linear
[see Fig.~\ref{bande_mouvement}(f)].

\subsection{Temporal fluctuations of the structure}

The previous measurements clearly show a displacement of the interface between the two bands in the gap. As the two bands
present different microscopic structures, we believe that this motion may also be detected on the
diffraction patterns. The patterns associated with the disordered 
texture are homogeneous rings, whereas six peaks modulate the ring at the onset of the layering transition.
These peaks are associated with the hexagonal long range order of the onions on planes oriented along the flow
\cite{Diat:95}. 
Let us recall that the diffraction patterns correspond to a measure of the 
structure {\it integrated} along the velocity gradient direction $\nabla\!v$.  
During the layering transition, the diffraction pattern evolves continuously from an homogeneous 
ring to six well-defined peaks \cite{Salmon:03_6}. 

\begin{figure}[ht!]
\begin{center}
\scalebox{1.0}{\includegraphics{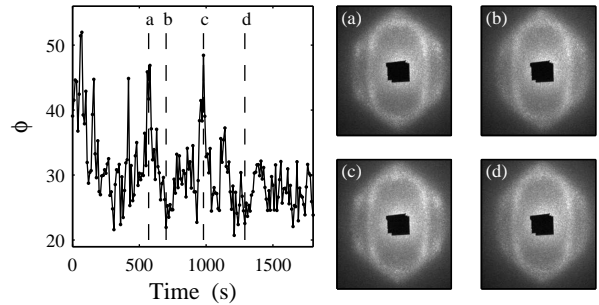}}
\end{center}
\caption{
Time series $\phi(t)$ (arbitrary unit) measured simultaneously to the data displayed in Figs.~\ref{global_rheology} 
and \ref{moyenne_6profils} at $\gammap=22.5$~s$^{-1}$ and $T=30^\circ$C. (a)--(d) Diffraction patterns 
obtained at the times indicated by the dotted lines.
}
\label{fluctu_phi}
\end{figure}

A series of images were taken every 10~s at $\gammap=22.5$~s$^{-1}$ and $T=30^\circ$C in the coexistence domain.
To enhance the signal-to-noise ratio, each image corresponds to an average over five successive 
pictures separated by 40~ms. 
From the intensity of the peaks relative to that of the ring, we can compute a {\it degree of organization} of the sample
\cite{Salmon:02}. 
Such a parameter $\phi$ is defined as the difference between the maximal intensity and the minimal 
intensity measured on the ring. 
$\phi$ is significanty larger than 0 when peaks modulate the ring, whereas $\phi\approx 0$ when rings are uniform.
Figure~\ref{fluctu_phi} presents the evolution of 
$\phi$ as a function of time for $\gammap=22.5$~s$^{-1}$.
First, note that $\phi(t)$ is always positive since peaks are always present on the ring:
the intensity modulation of the ring is due to the presence of a highly sheared band most probably
corresponding to the layered state of onions. Indeed, as shown in Fig.~\ref{moyenne_6profils}(b), the width $\delta$
of the ordered band is about $0.5$~mm at the considered $\gammap$.

However, the time series $\phi(t)$ fluctuates
with characteristic periods of 100--500~s. 
The amplitude of these fluctuations is rather small but still meaningful since some well-chosen
patterns [see Figs.~\ref{fluctu_phi}(a)--(d)] clearly show the variation of the contrast of the peaks on the ring.  
These structural fluctuations support the idea that the proportion of ordered onions relative to disordered onions
constantly changes, which is consistent with a motion of the interface between the bands.
Moreover, it is reasonable to assume that such an ordering/disordering 
process, leading to the displacement of the highly sheared band, 
involves time scales longer than the characteristic times of hydrodynamics. 
Such long time scales correspond to those observed in the dynamics of $\sigma(t)$ and
of $v(x_i,t)$. 

\section{Analysis using a simple mechanical approach \label{auasma}}

\subsection{Constitutive equations of the mechanical approach}

\begin{figure}[ht!]
\begin{center}
\scalebox{1.0}{\includegraphics{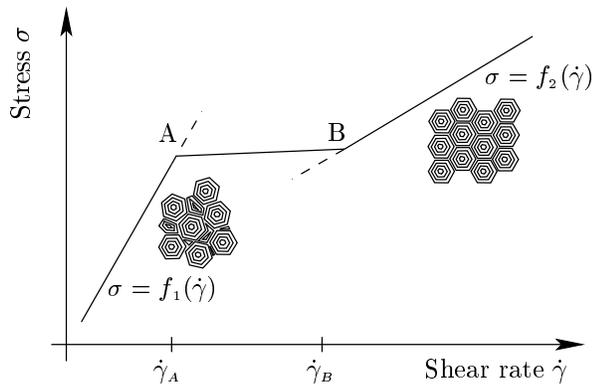}}
\end{center}
\caption{Flow curve of the layering transition. The contribution due to wall-slip are removed.
A--B indicates the coexistence domain between the two textures.}
\label{modele2}
\end{figure}

In the first part of this article \cite{Salmon:03_6}, we have shown that the time-averaged velocity profiles may be 
well fitted by a simple mechanical approach. 
Using the rheological data $\sigma$ vs. $\gammap$ together with the assumption that the interface 
between the bands always lies at the same local stress $\sigma^\star$, we were able to reproduce the velocity data.
Let us recall briefly such a mechanical approach.
The flow curve obtained from the rheological data where the contribution due to wall-slip are
removed is sketched in Fig.~\ref{modele2}. 
The shear rate plotted on this flow curve is
the effective shear rate in the bulk material once slip effects are removed. 
The homogeneous branches corresponding to the 
disordered and layered states, yield the local rheological behaviors
$\sigma = f_i(\gammap)$, where $i=1$ ($i=2$ resp.) denotes the disordered (layered resp.) texture.
 
Due to the the curvature of the Couette cell, the local stress $\sigma(r)$ varies in the gap as:
\begin{equation}
\sigma(r) = \frac{\Gamma}{2\pi H r^2} = \sigma_{{\scriptscriptstyle 1}}
\frac{\rot^2}{r^2} \,,
\label{stress_couette}
\end{equation}
where $r$ is the radial position in the gap, $\Gamma$ is the imposed torque and 
$\sigma_{{\scriptscriptstyle 1}}=\Gamma/(2 \pi H \rot^2)$ is the stress at the rotor. 
In our mechanical picture, the interface lies at a given stress $\sigma^\star$. The width $\delta$ of the highly
sheared band is thus given by:
\begin{equation}
\delta = \rot\left(\sqrt{\frac{\sigma_{\scriptscriptstyle 1}}{\sigma^\star}}-1\right)\,.
\label{hbande_couette}
\end{equation}
It is then straightforward to compute the theoretical profiles
from an arbitrary value of $\Gamma$ \cite{Salmon:03_6}.
If $\sigma(r)<\sigma^\star$ ($\sigma(r)>\sigma^\star$ resp.) everywhere in the gap, the flow is homogeneous
and composed of the disordered (layered resp.) state of onions.
In that case, the velocity profiles are  
given by the following integration of the rheological
behavior:
\begin{equation}
\frac{v(r)}{r} = \frac{v_{\scriptscriptstyle 2}}{\sta}+\int_{\sta}^{r}
\frac{\gammap(u)}{u}\,\hbox{d}u\,,
\label{int1}
\end{equation}     
where $\gammap(r)$ is found by solving:
\begin{align}
\sigma(r)= \frac{\Gamma}{2\pi H r^2}=f_i\left(\gammap(r)\right)\,, 
\label{int2}
\end{align}
and $i$ denotes the considered branch.
When there exists one particular position in the gap where $\sigma(r)=\sigma^\star$,
the flow displays two bands supporting different shear rates. To calculate the resulting velocity profile,
one should separate the previous integration between $[\sta;\rot+\delta]$ and $[\rot+\delta;\rot]$
(see Ref.~\cite{Salmon:03_6} for details).

From this set of equations, one can compute a theoretical profile $v(x)$ from an arbitrary value 
of $\Gamma$. It is then straigthforward to infer a global shear rate $\gammap$ indicated by the rheometer.
Such a procedure has been applied to our data in Ref.~\cite{Salmon:03_6}:
since Eqs.~(\ref{int1}) and (\ref{int2}) lead to satisfactory fits of both the flow curve and of the velocity profiles, 
the assumption that the interface lies at a given stress $\sigma^\star$, holds for the layering transition.  

\subsection{Temporal fluctuations of $\sigma^\star$}

One point is particularly important: for a given value of torque $\Gamma$, there exists only
one profile $v(x)$. Thus, under controlled shear rate, these equations predict that 
the stress $\sigma$ indicated by the rheometer
is well-defined and that the position of the interface is fixed. 
Moreover, one can easily check that if $\sigma^\star$ is fixed, very small variations of 
the applied shear rate cannot induce large fluctuations of the measured $\sigma$. Thus,
the experimental data displayed in Fig.~\ref{global_rheology} cannot be understood simply by these equations. 

\begin{figure}[ht!]
\begin{center}
\scalebox{1.0}{\includegraphics{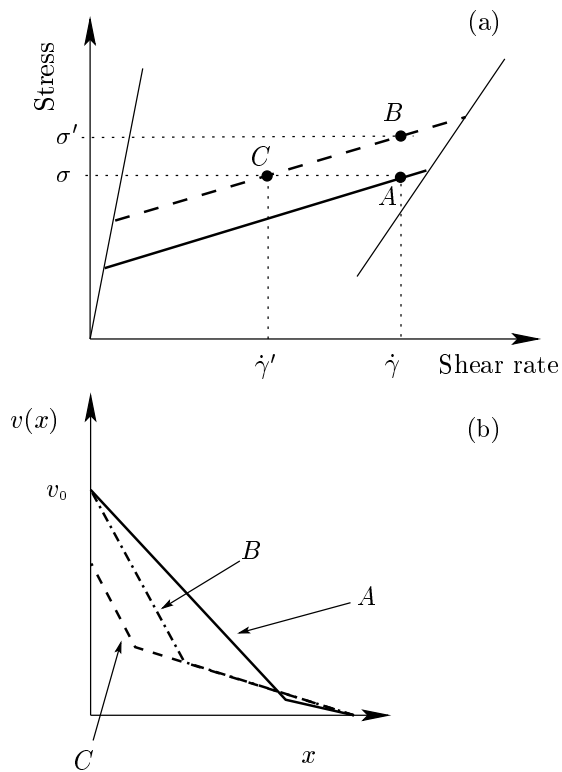}}
\end{center}
\caption{
(a) Schematic flow curve for the shear-banding scenario in the Couette geometry.
When the stress $\sigma^\star$ fluctuates, the position of the stress {\it plateau} 
changes from the continous line to the dotted line. Point A indicates the 
initial configuration of the flow field.
B: final configuration under controlled shear rate, the measured stress
has varied from $\sigma$ to $\sigma'$. C: final configuration under controlled stress,
the shear rate has varied from $\gammap$ to $\gammap'$.
(b) Corresponding velocity profiles. $v_{\scriptscriptstyle 0}$ is the rotor velocity. 
}
\label{sigtar}
\end{figure}

However, since Eqs.(\ref{int1}) and (\ref{int2}) describe nicely the time-averaged velocity profiles, we believe that
if a crucial parameter of the previous equations possesses its own hidden dynamics, the mechanical
approach may still help us understand the observed fluctuations. 
Since the interface between the two bands fluctuates in the gap and since its position 
is fixed by $\sigma^\star$ through Eq.~(\ref{hbande_couette}), we
can reasonably assume that some fluctuations of $\sigma^\star$ could explain our data. 
Figure~\ref{sigtar} presents a sketch of this scenario. Consider an initial state A in the 
shear-banding region. 
The flow profile presents two differently sheared bands [see Fig~\ref{sigtar}(b)]. 
Let us now assume that $\sigma^\star$ slightly increases. The position of the {\it stress plateau} thus changes,
as well as the position in the gap that corresponds to $\sigma^\star$. 
The interface does not lie on a stable position anymore.
Under controlled shear rate, the final stable state will lie at point B on the flow curve, and the global stress
will vary from $\sigma$ to $\sigma'$. This simple picture helps us understand why fluctuations 
of $\sigma$ could be observed without any significant fluctuations of $\gammap$ and with a motion of the interface in the gap. 

\begin{figure}[ht!]
\begin{center}
\scalebox{1.0}{\includegraphics{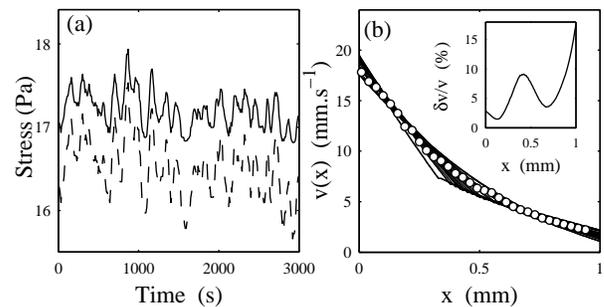}}
\end{center}
\caption{(a) The continuous line is the experimental time series $\sigma(t)$
measured at $T=30^\circ$C and at $\gammap=22.5$~s$^{-1}$. The dashed line is the 
time series $\sigma^\star(t)$ used to compute the theoretical velocity profiles in our
mechanical analysis (see text). 
(b) Corresponding experimental velocity profile ($\circ$). 
The bundle of continous lines corresponds to the theoretical velocity profiles 
computed from Eqs.~(\ref{int1}) and (\ref{int2}) at $\gammap=22.5$~s$^{-1}$ and with $\sigma^\star(t)$ displayed in (a).
Inset: amplitude of the theoretical temporal fluctuations inferred from the standard deviation of the theoretical 
profiles vs. position $x$.  
}
\label{model_fluctu}
\end{figure}

To proceed further into the analysis, one should compare such a scenario to the experimental data.
Since the global stress $\sigma$ fluctuates as $\sigma^\star$, we assume that the fluctuations 
of $\sigma^\star$ have the same shape as those of $\sigma(t)$.
Figure~\ref{model_fluctu}(a) displays a time series $\sigma^\star(t)$ inferred from a typical time series 
$\sigma(t)$ obtained under controlled shear rate at $\gammap=22.5$~s$^{-1}$.
The mean value of $\sigma^\star$ is set at 16.15~Pa, the value leading to the best fits of the experimental data
in Ref.~\cite{Salmon:03_6}.

\begin{figure}[ht!]
\begin{center}
\scalebox{1.0}{\includegraphics{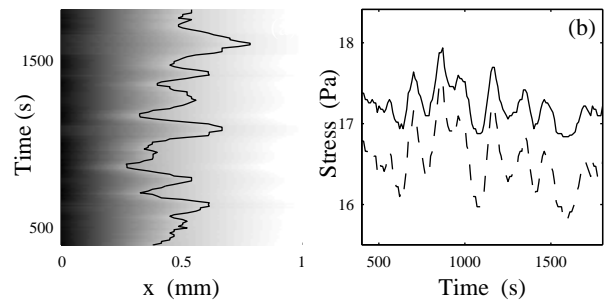}}
\end{center}
\caption{
(a) Time-space plot of the theoretical velocity profiles displayed in Fig.~\ref{model_fluctu}(b)
and calculated with the time series $\sigma^\star(t)$ presented in (b) (dashed line). 
The black line is the theoretical width $\delta$ of the highly sheared band
calculated using Eq.~(\ref{hbande_couette}).
(b) Corresponding time series $\sigma(t)$ (continuous line) and $\sigma^\star(t)$
(dashed line). 
}
\label{bande_fluctu}
\end{figure}

We have presented in Fig.~\ref{model_fluctu}(b), 	
the resulting theoretical velocity profiles computed from Eqs.~(\ref{int1}) and (\ref{int2}) with $\sigma^\star(t)$
displayed in Fig.~\ref{model_fluctu}(a). Figure~\ref{model_fluctu}(b) also presents the experimental time-averaged 
velocity profile obtained at $\gammap=22.5$~s$^{-1}$ for comparison.  
Figure~\ref{bande_fluctu} displays a time-space plot of the same theoretical velocity profiles.
The time series $\delta(t)$ of the position of the interface 
computed from Eq.~(\ref{hbande_couette}) has also been added on this time-space plot.

Our crude approach succeeds to describe many experimental points. (i) The fluctuations 
of $\sigma^\star$ induce a large fluctuating motion of the interface
(see Fig.~\ref{bande_fluctu}). (ii) The amplitude of
the fluctuations of the flow field is thus localized [see inset of Fig.~\ref{model_fluctu}(b)]: 
they reach a maximum of about 10\% at the position of the 
interface. (iii) A quantitative agreement may be found between the theoretical fluctuations 
and the measured data within this crude approach: fluctuations of $\sigma^\star$ by 2\%  
induce fluctuations of the measured $\sigma$ by 1\%, and fluctuations of the local velocities reaching 10\%
at the interface. Let us outline that fluctuations of the others parameters of the 
mechanical equations such as the viscosities of the two different textures will not be able 
to reproduce these dynamics.

\subsection{Validation under controlled shear stress}

To fully validate our mechanical approach, one should perform experiments under
controlled shear stress, and show that the fluctuations of $\sigma^\star$ may also induce a 
fluctuating motion of the interface and some variations of the measured $\gammap(t)$. 

However, for temperatures $T\geq T_c=27^\circ$C and under controlled stress, 
the lyotropic lamellar phase under study displays well-defined oscillating
behaviors of $\gammap(t)$ with a period of about 500~s and an amplitude reaching about 20~s$^{-1}$
in the vicinity of the transition.
Unfortunatly, our setup is not suitable to follow such dynamics of the flow field, since
the involved periods are of the order of minutes and the amplitudes are large. Using DLS velocimetry, 
it is thus rather difficult to determine
whether the shear rate oscillations observed for $T\geq T_c$ are due to an oscillating
motion of the interface. 
We thus decide to perform experiments for temperatures $T<T_c$, where 
the shear rate only displays fluctuating behaviors rather than sustained oscillations. 
Indeed, in this region of parameters, the amplitude of the dynamics
is rather small and the dynamics is slow enough to be captured by our setup.

\begin{figure}[ht!]
\begin{center}
\scalebox{1.0}{\includegraphics{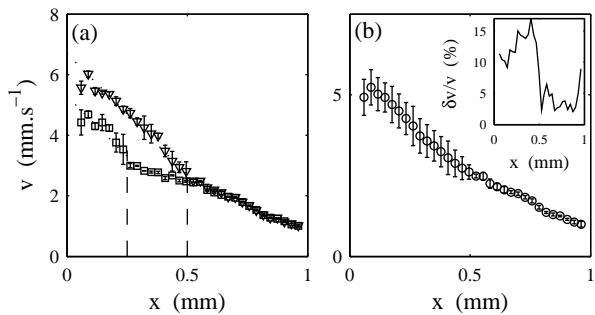}}
\end{center}
\caption{
(a) Two {\it averaged groups} of velocity profiles obtained at $\sigma=12.5$~Pa and $T=26^\circ$C.
The vertical dashed lines lie at $x=0.25$ and $x=0.50$~mm.
The dotted lines help to guide the eye in the highly sheared band.
(b) Velocity profile averaged over 14 consecutive measurements at the same stress.
The errorbars are the standard deviations of these estimates.
Inset: temporal fluctuation of the local velocity inferred from the standard deviation of the 
different measurements vs. $x$.}
\label{bande_mouvement_stress}
\end{figure}

Figure~\ref{bande_mouvement_stress}(a) presents two well chosen groups of velocity profiles measured
in the coexistence domain under controlled
$\sigma$ and at $T=26^\circ$C. The velocity profile averaged over 14 successive measurements
is shown in Fig.~\ref{bande_mouvement_stress}(b). These data clearly show a very large motion of the interface 
in the gap (from $x=0.25$ to 0.50~mm). In this experiment under controlled stress, the velocity
of the fluid near the rotor and thus the measured $\gammap$ vary a lot.
Following our crude mechanical approach, the situation is the one sketched in Fig.~\ref{sigtar}.
The flow field is initially stable at point A. A small variation of $\sigma^\star$ changes the position
of the plateau on this flow curve. Under controlled stress, the final stable configuration lies at point C: 
the shear rate has varied from $\gammap$ to $\gammap'$ and the interface has moved in the gap.   

\section{Conclusions, discussions, and perpectives}

The present work has provided an exhaustive study of a shear-induced
transition in a specific complex fluid: the {\it onion texture} of a lyotropic lamellar phase. 
Under controlled shear rate and above some critical value, onions get a long range hexagonal order into planes 
sliding onto each other. 
In the vicinity of this ordering transition, one observes a spatial organization  
of the flow field: a band corresponding to the ordered texture is nucleated near the rotor and coexists
with the disordered texture. As pointed out by our velocity measurements, these two
structures flow with different shear rates and the highly sheared band
invades the gap as $\gammap$ is increased. 
The proportion of the two differently sheared bands in the flow field is governed 
by the controlled $\gammap$ and by $\sigma^\star$, the local stress at the interface. 

Our experiments have also revealed the presence of large fluctuations of 
the flow field localized at the interface between the two bands.
Using DLS velocimetry coupled to rheology and structural measurements, we have shown that 
these temporal fluctuations are due to a fluctuating motion of the interface in the gap.
The characteristic times of these puzzling dynamics range between 100 and 1000~s and do not 
correspond to external noise.
Using the classical mechanical picture of shear-banding, we have demonstrated 
that the fluctuations of $\sigma^\star$ may explain the observed dynamics. Indeed, under controlled shear rate, 
small variations of $\sigma^\star$ may induce small variations of $\sigma$, but also large 
fluctuations of the interface between the bands and thus localized fluctuations of the flow field.  
Under controlled stress, we have also shown that for $T<T_c=27^\circ$C, fluctuations of $\sigma^\star$ may describe 
correctly the observed displacement of the interface.

This crude approach describes very nicely the dynamics of the fluctuations but raises a crucial
question: why is $\sigma^\star$ a fluctuating parameter? Clearly, this parameter does 
not vary because of some external mechanical vibrations. 
Following Lu's theoretical approach \cite{Lu:00}, this parameter depends on the mathematical form of the
model which describes the transition and its origin is due to the presence of gradient terms. 
In the phenomenological model of Ref.~\cite{Goveas:01} for instance, $\sigma^\star$ 
depends on the rate of change between the two structures of the differently sheared bands.
Our data suggest that a dynamical equation describing the evolution of $\sigma^\star$ 
is missing. 
Such an equation is difficult to derive since it must describe precisely the microscopic structure of the system.
Other theoretical approaches such as those of Ref.~\cite{Ajdari:98}, based on a dynamical equation for the motion of 
the interface, may also fully reproduce the observed fluctuations if other dynamical equations concerning the structure 
are added. Understanding the origin of the fluctuations of $\sigma^\star$ from a microscopic point of 
view remains a major challenge in the study of shear-banding.

Finally, another point is left unclear: above the critical temperature $T_c$ and under controlled stress, 
why do the dynamics
seem to become deterministic or even seem to display chaotic behaviors \cite{Salmon:02}?  
To answer such a crucial question, it is necessary to increase the temporal resolution of our setup. 
We plan to use a spectrum analyzer to reduce 
the acquisition time of the velocity measurements or to turn to other velocimetry methods 
such as ultrasonic techniques \cite{Manneville:03}. These experimental improvements will certainly help to 
get a more precise spatio-temporal picture of rheochaos in complex fluids.  

\begin{acknowledgments}
The authors are deeply grateful to D. Roux, L. B\'ecu, and C. Gay for many discussions and comments on this work.
The authors would like to thank the {\it Cellule Instrumentation} at CRPP for the realization of the heterodyne DLS setup.
We also wish to thank  A. Ajdari, A. Aradian, C. Barentin, L. Bocquet, L. Chevillard, 
F. Molino, G. Picard, L. White-Benon and C. Ybert for many helpful conversations. 
\end{acknowledgments}

\bibliographystyle{apsrev}

\end{document}